\documentclass[aps,showpacs,preprint,graphics]{revtex4}
\usepackage{amssymb}
\usepackage[dvips]{graphicx}
\usepackage{amsmath}
\usepackage{bm}
\usepackage{epsfig}
\usepackage{lscape}
\usepackage{rotating}
\usepackage{epsfig}

\begin{document}

\title{Gravitational waves and magnetic monopoles during inflation with Weitzenb\"{o}ck torsion.}

\author{ $^{1}$ Jes\'us Mart\'in Romero\thanks{E-mail address: jesusromero@conicet.gov.ar},$^{1,2}$ Mauricio Bellini \thanks{E-mail address: mbellini@mdp.edu.ar} and $^{3}$ Jos\'e Edgar Madriz Aguilar \thanks{E-mail address: madriz@mdp.edu.ar} }
\affiliation{ $^{1}$ Departamento de F\'isica, Facultad de Ciencias Exactas y Naturales\\
    Universidad Nacional de Mar del Plata, Funes 3350, C.P. 7600, Mar del Plata, Argentina\\
     $^{2}$ Instituto de Investigaciones F\'isicas de Mar del Plata (IFIMAR)\\
     Consejo Nacional de Investigaciones Cient\'ificas y T\'ecnicas (CONICET), Argentina.\\
     E-mail: jesusromero@conicet.gov.ar, mbellini@mdp.edu.ar\\
     $^{3}$ Departamento de Matem\'aticas, Centro Universitario de Ciencias Exactas e ingenier\'{i}as (CUCEI),
Universidad de Guadalajara (UdG), Av. Revoluci\'on 1500 S.R. 44430, Guadalajara, Jalisco, M\'exico.  \\
E-mail: madriz@mdp.edu.ar, edgar.madriz@red.cucei.udg.mx}

\begin{abstract}
We study the variational principle on a Hilbert-Einstein action in an extended
geometry with torsion taking into account non-trivial boundary conditions. We obtain an effective energy-momentum tensor that has its source in the torsion, which
represents the matter geometrically induced. We explore about the existence of magnetic monopoles and gravitational waves
in this torsional geometry. We conclude that the boundary terms can be identified as possible
sources for the cosmological constant and torsion as the source of magnetic monopoles. We examine an example in which gravitational waves
are produced during a de Sitter inflationary expansion of the universe.
\end{abstract}

\maketitle

\vskip .5cm Torsion, Weitzenb\"{o}ck Geometry, Magnetic Monopoles,
Gravitational Waves, Inflation.

\section{Introduction}

In the standard treatment of the variational principle over the
Hilbert-Einstein action (HE), when a manifold has a boundary
$\partial{\cal{M}}$, the action is supplemented by a boundary term
which is in general neglected\cite{HE}. However, this is not the only
manner to study this problem. As was recently demonstrated in\cite{2},
it is possible to include the flux around an hypersurface
that encloses a physical source without the inclusion of extra
terms in the HE action. In that paper was demonstrated that the
non-zero flux of the vector metric fluctuations through the closed
3D Gaussian-like hypersurface, is responsible for the
gauge-invariance of gravitational waves (GW). However, the torsional contributions were neglected
in that paper. In the present paper we extend this analysis on the
variational principle, but for an extended
geometry with torsion. We obtain an effective energy-momentum
tensor with sources only in torsion, which can be viewed as an
effective matter tensor in a Riemannian geometry. Such tensor
represents matter geometrically induced, but without extra dimensions.
In addition, we develop a new manner to obtain GW
on a torsional manifold taking into account nontrivial boundary
terms. The first contribution to GW with purely
torsional nature, was studied in Sect. III A, and the second one
based in the boundary term was studied in Sect. III B; in both
cases for a general torsion. Also, we present an example in
Weitzenb\"{o}ck geometry obtaining the expression for the magnetic
density monopoles and a gravitational wave equation for a Friedman-Robertson-Walker (FRW) in
Weitzeb\"{o}ck geometry. Finally, in Sect. VI, we develop some
final remarks.

\section{Variational Principle in Torsional Geometry.}\label{II}

We consider the variational principle
in presence of torsion for a HE-like action. We have studied
this fundamental problem in \cite{1}. Therefore, we shall
expose some results in present section without making a full description.
In \cite{1}, we have studied the contribution of the new terms which are not
present in a Riemannian geometry. The boundary term was studied in
\cite{2}, but emphasizing the role
of non-metricity. Now, we shall consider some gravitational action in an
extended geometry (i.e. a non-Riemannian manifold), without the presence of matter
\begin{eqnarray}\label{ac1}I=\frac{1}{2\kappa} \int_M d^4x \sqrt{- g}
R,\end{eqnarray}
in which $\kappa=8\pi\,G$, such that $G$ is the gravitational constant, and
\begin{eqnarray}
\label{RC} R^{m}_{\,\,\,lij}&=&\Gamma^{m}_{\,\,\,lj\,,i}-\Gamma^{m}_{\,\,\,li\,,j}+\Gamma^{n}_{lj}\Gamma^{m}_{\,\,\,ni}
-\Gamma^{n}_{\,\,\,li}\Gamma^{m}_{\,\,\,nj},\\
\label{Ric} R_{lj}&=&  R^{i}_{\,\,\,lij} ,
\end{eqnarray}
where $R=R_{nm}g^{nm} \label{R}$ is the scalar curvature.
We have employed the Einstein's convention over repeated indexes. The "$,$" represents a partial
derivative and all the indices run between $1$ and $4$. Furthermore,  $g_{ab}$ are the components of the metric tensor and $\sqrt{-g}$ is the volume of the non-Riemannian manifold. The eq. (\ref{RC}) defines the Riemann curvature tensor, the eq. (\ref{Ric}) give us the Ricci tensor and eq. (\ref{R}) is the scalar curvature. We denote with $\Gamma^{a}_{bc}$ an arbitrary affine connection, which
is defined according to
\begin{eqnarray}\label{cone1}
\nabla_{\overrightarrow{e}_a}\overrightarrow{e}_b=\Gamma^{n}_{ba}\overrightarrow{e}_n,
\end{eqnarray}
where $\nabla_{\overrightarrow{e}_a}$ denotes the derivative in $a$-direction of the tangent space $\left\{\overrightarrow{e}_b\right\}$.
Here, the up arrow means that the tangent space in the position representation is described by partial derivatives with respect to contravariant coordinates:
$\left\{\overrightarrow{e}_b\right\} \equiv \left\{\frac{\partial}{\partial x^b}\right\}$, and the down arrow means that the cotangent space is generated by $\underrightarrow{e}^b \equiv \left\{dx^b\right\}$, such that $\underrightarrow{e}^b(\overrightarrow{e}_a)=\delta^b_{\,\,a}$. We wont consider any particular symmetry in
the connections. Now we shall make the variation of the action in (\ref{ac1}): $\delta I=0$. Here we must take into account
that the scalar $R$ in (\ref{R}) is related to the connection in (\ref{cone1}), which is an abstract connection which is in
general non-Riemannian, but fulfils the expression
\begin{eqnarray}\label{cone2}
\Gamma^{n}_{mr}=\left\{^{n}_{mr}\right\}+K^{n}_{\,\,\,mr},
\end{eqnarray}
with $\{^{n}_{mr}\}$ the second kind Christoffel symbols representing the Riemannian or Levi-Civita connections, and
$K^{n}_{mr}$ the contortion tensor, which in absence of
non-metricity is entirely torsional according to
\begin{eqnarray}\label{k}
K^{a}_{\,\,\,bc}=-\frac{g^{na}}{2}\{T^{s}_{\,\,\,cn}\,g_{bs}+T^{s}_{\,\,\,bn}\,g_{sc}-T^{s}_{\,\,\,cb}\,g_{sn}\},
\end{eqnarray}
with the torsion tensor defined by
\begin{eqnarray}\label{tor}
T^{n}_{mr}&=&\Gamma^{n}_{rm}-\Gamma^{n}_{mr},
\end{eqnarray}
which is a valid expression in a coordinate basis of the four
dimensional tangent space to the space-time manifold (TM4). In
present work we impose the non-metricity free condition
\begin{eqnarray}\label{nom}
N_{nmr}&=&g_{nm\,;r}=0,
\end{eqnarray}
for an analysis of such contribution to the GW the reader can see \cite{1}.
The variation of the Ricci must be related to the variation of the connections obtaining a generalised Palatini
identity for torsional geometry
\begin{eqnarray}\label{gabdRab}
g^{mr}\,\delta
R_{mr}={W^{n}}_{\,\,;n}-\frac{1}{2}g^{mr}(\delta\Gamma^{n}_{\,\,\,pr}\,T^{p}_{\,\,\,mn}+\delta\Gamma^{n}_{\,\,\,pm}\,T^{p}_{\,\,\,rn}),
\end{eqnarray}
with
\begin{eqnarray}\label{w1}
W^{n}_{\,\,\,mr}=\,\delta\Gamma^{n}_{\,\,\,mr}-\,\delta\Gamma^{k}_{\,\,\,kr}\,\delta^{n}_{\,\,m}.
\end{eqnarray}
where $W^{n}=g^{mr}W^{n}_{\,\,\,mr}$. With the use of eq. (\ref{gabdRab}) in the variation of the action we obtain
\begin{eqnarray}\label{var2}
\delta I&=& \int_M d^4x\,\sqrt{-g}\, (R_{ab}-\frac{1}{2}R\,g_{ab})\,\delta g^{ab} \nonumber  \\ &+& \int_{\partial M} \,W^n\, d\Sigma_n \nonumber \\
&-&\frac{1}{2}\int_M d^4x\,\sqrt{-g}(\delta\Gamma^{n}_{pr}\,T^{p}_{mn}+\delta\Gamma^{n}_{pm}\,T^{p}_{rn})g^{mr}.
\end{eqnarray}
In the first integral we recognize the Einstein tensor for the
torsional connection. The second one is due to the boundary term. The third integral is completely originated by the torsion. This is
a non-Riemannian contribution.

To finalize this section we must present the explicit form of the
$W^{n}_{\,\,\,mr}=W^{n}_{\,\,\,(mr)}+W^{n}_{\,\,\,[mr]}$ tensor, where the symmetric and antisymmetric contributions are, respectively given by
\begin{eqnarray}
\label{w2}
W^{n}_{\,\,\,(mr)} & = &\left[\frac{g^{kn}}{2}\left\{\delta g_{mk\,,r}+\delta g_{kr\,,m}-\delta g_{rm\,,k}-T^{t}_{\,\,\,rk}\,\delta g_{mt}\right.\right.\nonumber \\
&-& \left.T^{t}_{\,\,\,mk}\,\delta g_{tr}\right\} - \frac{\delta g^{kn}}{2}\left\{g_{mk\,,r}+ g_{kr\,,m} \right. \nonumber \\
&-& \left. g_{rm\,,k}-T^{t}_{\,\,\,rk} \, g_{mt}-T^{t}_{\,\,\,mk}\,g_{tr}\right\}- \frac{g^{kl}}{4}\left(\delta g_{kl\,,r}\,\delta^{n}_{\,\,m}\right. \nonumber \\
&+& \left.\left.\delta g_{kl\,,m}\,\delta^n_{\,\,r}\right)+\frac{\delta g^{kl}}{4}\left( g_{kl\,,r}\,\delta^n_{\,\,m}+g_{kl\,,m}\,\delta^n_{\,\,r} \right) \right], \\
W^{n}_{\,\,\,[mr]}&=& \left[\frac{g^{kn}}{2}T^{t}_{\,\,\,rm}\,\delta g_{tk}-\frac{g^{kn}}{2}T^{t}_{\,\,\,rm}\,g_{tk}-\frac{g^{kl}}{4} \left(\delta g_{kl\,,r}\,\delta^n_m \right.\right. \nonumber \\
&-& \left.\left.\delta g_{kl\,,m}\,\delta^n_{\,\,r}\right) + \frac{\delta g^{kl}}{4}(g_{kl\,,r}\,\delta^n_{\,\,m}-g_{kl\,,m}\,\delta^n_{\,\,r})\right],
\end{eqnarray}
such that $W^n=W^n_{\,\,\,(mr)}\,g^{mr}$.

\section{Physics of the Torsional Geometry and 4D Induced Matter.}\label{III}

In presence of torsion, but zero non-metricity, the variation of the
action takes the form
\begin{eqnarray}\label{var4}
\delta I&=& \int_M d^4x\,\sqrt{-g}\, \left[R_{ab}-\frac{1}{2}R\,g_{ab}-\frac{1}{2}\,L_{(ab)}\right]\,\delta g^{ab}\nonumber \\
&+&\int_M d^4x\,\sqrt{-g}\,W^n_{\,;\,n}.
\end{eqnarray}
with
\begin{eqnarray}\label{lsd}
L_{(sd)}&=&\left\{\Delta^p_{\,\,\,mrsd}\,K^n_{\,\,\,pn}-\Delta^p_{\,\,\,nrsd}\,K^n_{\,\,\,pm}-\Delta^p_{\,\,\,nmsd}\,K^n_{\,\,\,pr}\right.\nonumber \\
&+& \left.\Delta^n_{\,\,\,npsd}\,(K^p_{\,\,\,rm}+K^p_{\,\,\,mr})\right\}
\,g^{mr}.
\end{eqnarray}
Furthermore
\begin{eqnarray}\label{triangulo}
\Delta^p_{\,\,\,mrsd}&=&\frac{g^{pk}}{2}\{-{(g_{ms}\,g_{kd})}_{,r}-{(g_{ks}\,g_{rd})}_{,m}+{(g_{ms}\,g_{rd})}_{,k}\nonumber  \\ &+& T^l_{\,\,\,rk}\,g_{ms}\,g_{ld}+T^l_{\,\,\,mk}\,g_{ls}\,g_{rd}-T^l_{\,\,\,rm}\,g_{ls}\,g_{rd}\}\nonumber \\ &-& \frac{1}{2}(-T^l_{\,\,\,rd}\,g_{ml}-T^l_{\,\,\,md}\,g_{lr}+T^l_{\,\,\,rm}\,g_{ld})\,\delta^p_{\,\,s}.
\end{eqnarray}
The first integral in (\ref{var4}) includes the extended Einstein
tensor with the torsional (Weitzenb\"ock) contribution,  and the second one includes the boundary contribution.
We have obtained the expression (\ref{var4}) in absence of matter. We can distinguish two possible cases.
\begin{enumerate}
\item\label{ca1} The first case
describes infinity manifolds and there are no boundary contributions: $W^n_{\,;n} =0$, so that the first integrand in
(\ref{var4}) is null:
\begin{equation}\label{c1}
R_{ab}-\frac{1}{2}R\,g_{ab}=\frac{1}{2}\,L_{(ab)}.
\end{equation}
After some algebraic manipulation of $L_{(ab)}$, the last
assumption leads to a wave equation originated in the presence of
torsion:
\begin{eqnarray}\label{onda}
\Box \,(\Delta^n_{prsd}\,\delta g^{sd}\,T^p_{mn})=0,
\end{eqnarray}
where $\Delta^n_{prsd}$ is given by (\ref{triangulo}).\\
\item\label{ca2}
The second case describes finite manifolds so that the boundary contributions are significative: $W^n_{\,;n} \neq 0$.
In that case the first integrand must be nonzero in order to $\delta I=0$:
\begin{equation}\label{c2}
\delta g^{ab} \left[R_{ab}-\frac{1}{2}R\,g_{ab}-\frac{1}{2}\,L_{(ab)}\right]+W^n_{\,;\,n}=0.
\end{equation}
In the relativistic formalism without boundary conditions (i.e., when $W^n_{\,;\,n}=0$), the cosmological constant can be added to the Einstein equations
as an integration constant. Therefore, in order to recover the dynamics in absence of boundary conditions, it is reasonable to make the
interpretation that the second integral can be related to the cosmological constant as
\begin{equation}\label{tn2}
W^{n}_{\,\,;\,n}=\Lambda(x)\,g_{ab}\delta g^{ab}.
\end{equation}
Here, $\Lambda (x)$ is a function of the proper time on the
torsional manifold, and play the role of a dynamical
cosmological constant with source in the boundary term.
The identification in (\ref{c1}) defines geometrically the physical sources
originated in the torsion, without cosmological constant.

With the help of (\ref{cone2}) and the contortion tensor (\ref{k}) in the
equation (\ref{c2}), we recover the effective equation for the
Riemannian part of the Einstein tensor
\begin{eqnarray}\label{ein2}
^{(R)}R_{ab}-\frac{1}{2}\,^{(R)}R\,g_{ab}+\Lambda g_{ab}=kT_{ab},
\end{eqnarray}
where the supra-index  "$^{(R)}$" in $^{(R)}R_{ab}$ and $^{(R)}R$, indicates
that these objects are Riemannian and constructed entirely with
the Levi-Civita connection. The tensor $T_{ab}$ in (\ref{ein2}) is an effective and geometrically induced
energy-momentum tensor over the Riemannian space-time, which is
associated with the variation over the Weitzenb\"ock manifold
\begin{eqnarray}\label{Tab}
k \,T_{ab}&=&-K^{n}_{\,\,\,nb\,|a}+K^{n}_{\,\,\,ab\,|n}-K^{l}_{\,\,\,nb} K^{n}_{\,\,\,la}+K^{l}_{\,\,\,ab}K^{n}_{\,\,\,lm}\nonumber  \\ &+& \frac{g^{sd}}{2}\left(K^{n}_{\,\,\,nd\,|s}-K^{n}_{\,\,\,sd\,|n}+K^{l}_{\,\,\,nd}K^{n}_{\,\,\,ls}\right. \nonumber \\
&-& \left.K^{l}_{\,\,\,sd}K^{n}_{\,\,\,lm}\right.)\,g_{ab},
\end{eqnarray}
which is of torsional nature in a clear way. We see, from last
expression, that the generalized Einstein tensor and the Riemannian
one, must be related according to
\begin{eqnarray}\label{ein}
G_{ab}&=&\,^{(R)}G_{ab}+K^{n}_{nb\,|a}-K^{n}_{ab\,|n}+K^{l}_{nb}K^{n}_{la}-K^{l}_{ab}K^{n}_{lm}\nonumber  \\ &-& \frac{g^{sd}}{2}\left( K^{n}_{\,\,\,nd\,|s}-K^{n}_{\,\,\,sd\,|n}+K^{l}_{\,\,\,nd}K^{n}_{\,\,\,ls}\right. \nonumber \\
&-& \left. K^{l}_{\,\,\,sd}K^{n}_{\,\,\,lm}\right)\,g_{ab}.
\end{eqnarray}
We must remark that $K^{n}_{nb\,|a}$ denotes the covariant
derivative of the torsional contorsion tensor according to the
derivative operator of the Riemannian geometry in which the
connection is the Christoffel symbol. From (\ref{ein2}) we
obtain the expression from its symmetric part and a consistence
equation for the anti-symmetric contribution. In presence of matter, which
is related to the lagrangian density $\mathcal{L}$ with an
energy-momentum tensor $^{(m)}T_{ab}$, we obtain
\begin{eqnarray}
G_{(ab)}+\frac{1}{2}\Lambda\,g_{ab}=k\,^{(m)}T_{(ab)}\,&\rightarrow&\,^{(R)}G_{(ab)}
+\frac{1}{2}\Lambda\,g_{ab}=k\,\left[^{(m)}T_{(ab)}+T_{(ab)}\right],\label{ein4} \\
G_{[ab]}=R_{[ab]}=k\,^{(m)}T_{[ab]}\,&\rightarrow&\,^{(R)}G_{[ab]}=0=k\,\left[^{(m)}T_{[ab]}+T_{[ab]}\right]. \label{ein5}
\end{eqnarray}
Equations in the l.h.s. of (\ref{ein4}) and (\ref{ein5}) before
the arrows are Einstein-Cartan-like eqs. We assume a symmetric
metric tensor $g_{ab}=g_{(ab)}$. The tensor $T_{ab}$ is induced from
torsional geometry in the Weitzenb\"ock representation and is adequate to describe matter with spin or magnetic
monopoles. However, $^{(m)}T_{ab}$ must be proposed from an additional
term in the action. In this work we propose an empty torsional geometry with
$^{(m)}T_{ab}=0$, then the r.h.s. of the equation (\ref{ein4}) can be reduced to
\begin{equation}\label{ein3}
^{(R)}G_{(ab)}+\frac{1}{2}\Lambda\,g_{ab} = k\,T_{(ab)},
\end{equation}
and the r.h.s. of (\ref{ein5}) is
reduced to
\begin{equation}\label{ein6}
0 = T_{[ab]}.
\end{equation}
Notice that the equation (\ref{ein3}) corresponds to the usual Riemannian Einstein
equations, and (\ref{ein6}) are the conditions obtained from the
anti-symmetric part of the Einstein torsional tensor.
\end{enumerate}

\section{Weitzenb\"{o}ck geometry}\label{IV}

The Weitzenb\"{o}ck geometry could be formulated from the vierbein,
which are coefficients that express the relation between two different basis of TM4,
$\{\overrightarrow{E}_A\}$ and $\{\overrightarrow{e}_a\}$
\begin{eqnarray}\label{vier}
\overrightarrow{E}_A=e_A^a\,\overrightarrow{e}_a,\,\,\,\,\,\,\overrightarrow{e}_a=\bar{e}_a^A\,\overrightarrow{E}_A,
\end{eqnarray}
with the properties
\begin{eqnarray}\label{vier2 }
e_A^a\,\bar{e}^A_b=\delta^a_b,\,\,\,\,\,e_B^b\,\bar{e}^A_b=\delta^A_B,
\end{eqnarray}
such that the covariant derivative of basis elements $\bar{e}^A_{b}$ to be zero: $\bar{e}^A_{b;C}=0$.
If we take into account the equations (\ref{vier}) and (\ref{vier2 }), one can show that
the components of some arbitrary tensor $T$ in
${\mathcal{T}^p}_m(M)$, transforms according to
\begin{eqnarray}\label{trafo}
{T^{a_1\,...\,a_p}}_{b_1\,...\,b_m}=e^{a_1}_{A_1}\,...\,e^{a_p}_{A_p}\,\bar{e}^{B_1}_{b_1}\,...\,\bar{e}^{B_m}_{b_m}{T^{A_1\,...\,A_p}}_{B_1\,...\,B_m}.
\end{eqnarray}
In this framework it is possible to define the Weitzenb\"{o}ck connection
\begin{eqnarray}\label{conewei}
^{(We)}\Gamma^a_{bc}=e^a_N\overrightarrow{e}_c(\bar{e}^N_b),
\end{eqnarray}
so that
\begin{eqnarray}\label{conewei1}
^{(We)}\Gamma^A_{BC}=0,
\end{eqnarray}
and therefore $^{(We)}R^A_{BCD}=0$. In the usual Weitzenb\"{o}ck basis is proposed that the basis $\{\overrightarrow{e}_a\}$ to be a coordinate basis of TM4, with certain metric characterized by $g_{ab}$, which is of interest. The basis $\{\overrightarrow{E}_A\}$ can be non-coordinate but must be chosen in the form that the metric tensor expressed in such basis to be characterized by $\eta_{AB}$. In this context the Weitzenb\"{o}ck torsion is
\begin{eqnarray}\label{tor1}
^{(We)}T^a_{bc}=e^a_A\,\bar{e}^B_b\,\bar{e}^C_c\,^{(We)}T^A_{BC}=e^a_A\,\bar{e}^B_b\,\bar{e}^C_c\,C^A_{BC},
\end{eqnarray}
where $C^A_{BC}$ are the structure coefficients of the basis $\{\overrightarrow{E}_A\}$. The structure must be taken into account in the equations (\ref{RC}),
(\ref{Ric}), (\ref{R}) and (\ref{tor}). The structure coefficients $\{\overrightarrow{E}_A\}$ are
defined by
\begin{eqnarray}\label{estruc}
[\overrightarrow{E}_B,\overrightarrow{E}_A]=C^C_{AB}\,\overrightarrow{E}_C.
\end{eqnarray}
The non-metricity related to the Weitzenb\"{o}ck connection is
\begin{eqnarray}\label{nomewe}
^{(We)}N_{abc}=\bar{e}^A_a\,\bar{e}^B_b\,\bar{e}^C_c\,^{(We)}N_{ABC}=\bar{e}^A_a\,\bar{e}^B_b\,\bar{e}^C_c\,\eta_{AB\,,C}.
\end{eqnarray}
In order to obtain (\ref{tor1}) and (\ref{nomewe}), we have used (\ref{conewei1}). Usually, we choose $\{\overrightarrow{E}_A\}$ in order to it
be an orthonormal basis, such that $\eta_{AB}=-1,0,+1$. Therefore one obtains $\eta_{AB\,,C}=0$, and the equation (\ref{nomewe}) becomes null:
\begin{eqnarray}\label{nomewe2}
^{(We)}N_{ABC}=0.
\end{eqnarray}
The usual Weitzenb\"{o}ck geometry is a torsional geometry with
zero non-metricity. Such elements characterize the Weitzenb\"{o}ck
connection (\ref{cone2}) with a contortion tensor, which is due as a function of the Weitzenb\"{o}ck torsion in (\ref{tor1}).
The zero non-metricty condition must be removed only if the elements of the basis $\{\overrightarrow{E}_A\}$ are chosen such that its
inner product is different that a constant.

\subsection{Magnetic Monopoles with Weitzenb\"{o}ck geometry.}

With a standard gravito-electromagnetic action for the extended geometry
we must obtain an extension of the Maxwell equations:
\begin{eqnarray}
*d(F)=\,^{(m)}J,\label{max} \\
*d(*F)=\,^{(e)}J.\label{min}
\end{eqnarray}
where $F$ is the extended Faraday $2$-form, $d$ is an exterior covariant derivative and $*$ is the
adjunction operation. The source term $^{(m)}J$ is a cotangent
vector of the magnetic current and $^{(e)}J$ is the electric current. In both terms the zero component $^{(*)}J_0=\rho_{*}$ is the
corresponding Hodge-dual of the charge density. The expression (\ref{max}) is different of the usual one $d(F)=0$, because
the torsion of the geometry is the responsible for a non zero current of magnetic monopoles. The adjunction is taken into account in order to
match correctly the vectorial order of both sides of the magnetic side in the equation (\ref{max}). The expression in (\ref{min}) has the usual
appearance of the Maxwell ones, but for gravito-electromagnetic currents produced by a torsional geometry.

A $p$-form is an anti-symmetric tensorial object $W$ of order $p$
\begin{eqnarray}\nonumber
W=\frac{1}{p!}\,w_{i_1\,...\,i_p}\,\underrightarrow{e}^{i_1}\wedge...\wedge\underrightarrow{e}^{i_p},
\end{eqnarray}
in which the wedge product is the anti-symmetrization of the tensor product. The exterior covariant derivative associated with
certain covariant derivative denoted by ($;$), is defined by
\begin{eqnarray}\nonumber
d(W)=\frac{1}{p!}\,w_{i_1\,...\,i_p\,;k}\,\underrightarrow{e}^k \wedge \underrightarrow{e}^{i_1}\wedge...\wedge\underrightarrow{e}^{i_p}.
\end{eqnarray}
The adjunction operation in a manifold of dimension $m$, is
\begin{eqnarray}\nonumber *W=\frac{\sqrt{|g|}}{(m-p)!\,p!}\varepsilon_{j_1\,...\,j_pi_{p+1}\,...\,i_n}w^{j_1\,...\,j_p}\,&\underbrace{\underrightarrow{e}^{i_{p+1}}\wedge...\wedge\underrightarrow{e}^{i_m}}&,\nonumber \\
&m-p&
\end{eqnarray}
which takes into account a $p$-form and gives us a $(m-p)$-form. The Einstein-Faraday $2$-form is defined from the exterior covariant derivative
of the $1$-form, which is the cotangent version of the tetra-vector
$A=(\varphi,\overrightarrow{A})$. Here, $\overrightarrow{A}$ is the
usual $3$-vector potential
\begin{eqnarray}\nonumber
F=d(A).
\end{eqnarray}
If the connection is symmetric, then $d(F)=d(d(A))=0$, which implies the absence of magnetic monopoles.

In the present work we are dealing with a torsional Weitzenb\"{o}ck geometry, so that we must apply the equation (\ref{max}) for the corresponding connections.
Then, the $0$-component of the magnetic current will be
\begin{eqnarray}\label{rocero}
[*d(F)]_0=\rho_m=-3{^{(We)}T^D_{21}\overrightarrow{E}_D(A_3)+\,^{(We)}T^D_{13}\overrightarrow{E}_D(A_2)+\,^{(We)}T^D_{32}\overrightarrow{E}_D(A_1)},
\end{eqnarray}
which implies that $\rho_m=\bar{e}^A_{a=0}\,^{(m)}J_A$, in the $\vec{e}_a$ basis.

\section{Example: Magnetic monopoles and GW from torsion in a FRW expansion}\label{V}

In this section we shall address the usual Weitzenb\"{o}ck scenario as
described in Sect. (\ref{IV}). In particular, we shall study the case (\ref{ca1}), where is absent the cosmological constant and the
boundary conditions are trivial. We shall start with an
ortho-normalized Lorentzian metric related to a non-coordinate basis: $\{\overrightarrow{E}_A\}=\{\partial_t, \,a(t)\, \partial_a \}$,
with the structure coefficients $C^A_{BC}$
\begin{eqnarray}\label{estruc1}
C^1_{10}=C^2_{20}=C^3_{30}=\frac{\dot{a}(t)}{a(t)},
\end{eqnarray}
and its counterpart with lower changed indices. The equation (\ref{estruc1}) must be used with (\ref{tor1}) in order to
obtain the Weitzenb\"{o}ck torsion. The non-zero vierbein are
\begin{eqnarray}\label{vier3}
e^{a=0}_{A=0}=1,\,e^i_I=a(t),\,\,\,\,\,\,\,\,\,\,\bar{e}^{A=0}_{a=0}=1,\,\bar{e}^I_i=a(t)^{-1},
\end{eqnarray}
which are only valid in the case in which $i=I$. Here, $I$ runs over
the three space indices of the basis $\vec{E}_A$, and $i$ runs over
the three space indices of the basis $\vec{e}_{a}$. This one is a
coordinate basis given by $\{\overrightarrow{e}_a\}=\{\partial_t,\partial_a\}$. The metric tensor $g_{ab}$, for the basis $\vec{e}_{a}$, takes
the form
\begin{eqnarray}\label{frw}[g]_{ab}=\left(%
\begin{array}{cccc}
  1 & 0 & 0 & 0 \\
  0 & -a^2(t) & 0 & 0 \\
  0 & 0 & -a^2(t) & 0 \\
  0 & 0 & 0 & -a^2(t) \\
\end{array}%
\right),
\end{eqnarray}
which describes an isotropic an homogenous spatially flat, FRW expanding universe with scale factor $a(t)$. If we use the (\ref{estruc1}) and (\ref{vier3}) in (\ref{tor1}),
we obtain that the non-zero components of the Weitzenb\"{o}ck torsion are $^{(We)}T^{(i)}_{(i)\,0}=e^{(i)}_A
e_{(i)}^B e^C_{c=0}\,^{(We)}T^A_{BC}=e^{(i)}_A e_{(i)}^B
e^C_{c=0}\,^{(We)}C^A_{BC}=\,^{(We)}C^{A=(i)}_{B=(i)\,\,C=0}=\frac{\dot{a}(t)}{a(t)}$. Therefore the non-zero torsion components will be
\begin{eqnarray}\label{tor2}
^{(We)}T^{(i)}_{(i)\,0}=\frac{\dot{a}(t)}{a(t)},
\end{eqnarray}
where the indices between parenthesis indicate that such indices are equal and the Einstein notation for repeated indexes is not used.
Using the equation (\ref{tor2}) in (\ref{rocero}), we obtain that the density of magnetic monopoles is
\begin{eqnarray}\label{rocero1}
\rho_m=0.
\end{eqnarray}
This implies a non-trivial absence of magnetic monopoles due to the absence of torsion of spatial nature.
On the other hand, we must notice that the wave equation
(\ref{onda}) is used in the present case. The expression (\ref{onda}) is simplified
because some elements of the torsion are null. From the equation
(\ref{tor2}),we obtain that (\ref{onda}) is reduced to
\begin{eqnarray}\label{onda1}
\Box\,\left[\frac{\dot{a}(t)}{a(t)}\left(\delta g^{IJ}-\delta
g^{I0}\right)\right]&=&0, \label{onda1} \\  \Box \, \delta
g^{00}&=&0. \label{on}
\end{eqnarray}
Notice that the expression (\ref{on}) is equivalent to a gauge choice.

\subsection{GR in a de Sitter inflation}

For the particular case in which $a(t)=e^{\,H_0\,t}$, we
see that $\frac{\dot{a}(t)}{a(t)}=H_0$. Thus the expression (\ref{onda1}) have particular solutions in the equation
\begin{eqnarray}\label{onda2}
\Box\,\delta g^{AB}=0.
\end{eqnarray}
This case is important because describes a de Sitter inflationary expansion with an equation of state: $P/\rho =-1$. Here, $P$ is the isotropic pressure and
$\rho$ is the energy density of the universe during the expansion. During a de Sitter expansion $\rho$ remains constant.

On the other hand the equation (\ref{onda2}) has the solution in the form of a free plane wave for
Weitzenb\"{o}ck derivative operators. This is a simple wave equation
expressed in terms of the basis $\vec{E}_A$ which is related to null connections. The solution of
(\ref{onda2}) admits an expansion of the form
\begin{eqnarray}\label{gww}
\delta g_{AB}(t,\vec{x}) &=& \frac{1}{(2\pi)^{3/2}}\,\sum_{M=+,\times} \,\int d^3 k \,\, e^M_{AB}(\hat{z})\nonumber \\
&\times & \left[ A_k \, e^{i\vec{k}.\hat{z}}  \, \chi_k(t) +
A^{\dagger}_k \, e^{-i\vec{k}.\hat{z}}  \, \chi^*_k(t)\right],
\end{eqnarray}
where index $M=+, \times$ denote the Transverse-Traceless (TT) polarizations
$+,\times$, on the plane orthogonal to $\vec{k}$, and $e^M_{AB}$ are
the components of the polarization tensor, such that
\begin{equation}\label{pol}
e^M_{AB} \,\bar{e}^{AB}_{M'} = \delta^M_{M'}.
\end{equation}
For the case in which the
scale factor is $a(t)=e^{H_0\, t}$, and the Hubble parameter is a
constant $H_0$, the equation of motion for the modes
$\chi_k(t)$ is
\begin{equation}\label{chi}
\ddot\chi_k(t) + 3 H_0 \dot\chi_k(t) + \left[\frac{k^2}{e^{2H_0 t}}\right] \chi_k(t)=0.
\end{equation}
The general solution for the modes in (\ref{chi}) can be written in terms of the first and second kind Hankel functions ${\cal H}^{(1,2)}_{\nu}\left[\frac{k e^{-H_0t}}{H_0}\right]$
\begin{eqnarray}\label{soluc}
\chi_{k}(t) =\,e^{-\frac{3}{2}H_0t}\left\{A\,{\cal H}^{(1)}_{3/2}\left[\frac{k e^{-H_0t}}{H_0}\right]+B\,{\cal H}^{(2)}_{3/2}\left[\frac{k e^{-H_0t}}{H_0}\right]\right\}.
\end{eqnarray}
Since $g^{AB}g_{AB} =4$, it is easy to prove that $\delta \left(g^{AB}g_{AB}\right)=0$, so that we obtain
\begin{equation}\label{ccon}
g_{AB}\,\delta g^{AB}\,+\,\delta g_{AB}\,
g^{AB}\,=0.
\end{equation}
Since the dynamics of $\delta{g}^{AB}$ is described by the linear differential equation (\ref{onda2}), which describes a wave dynamics on a background curved spacetime, $\delta{g}^{AB}$ can be written as a Fourier expansion of a tensor field
\begin{eqnarray}\label{gww1}
\delta g^{AB}(t,\vec{x}) &=& \frac{1}{(2\pi)^{3/2}}\,\sum_{M=+,\times} \,\int d^3 k \,\, \bar{e}_M^{AB}(\hat{z})\nonumber \\
&\times & \left[ \widetilde{A}_k \, e^{i\vec{k}.\hat{z}}  \,
\widetilde{\chi}_k(t) + \widetilde{A}^{\dagger}_k \,
e^{-i\vec{k}.\hat{z}}  \, \widetilde{\chi}^*_k(t)\right].
\end{eqnarray}
In order to be fulfilled the condition (\ref{ccon}), we need require that the
annihilation and creation operators comply with
\begin{eqnarray}\label{conmu}
A_k\,\chi_k\,+\,\widetilde{A}_k\,\widetilde{\chi}_k=0,\\A_k^{\dag}\,\chi_k^{*}\,+\,\widetilde{A}_k^{\dag}\,\widetilde{\chi}_k^{*}=0.
\end{eqnarray}
This implies that the field $\delta g^{AB}(t,\vec{x}) $ must be expanded in terms of the coefficients $A_{k}$ and $A_{k}^{\dagger}$, as
\begin{eqnarray}\label{gww2}
\delta g^{AB}(t,\vec{x}) &=& \frac{-1}{(2\pi)^{3/2}}\,\sum_{M=+,\times} \,\int d^3 k \,\, \bar{e}_M^{AB}(\hat{z})\nonumber \\
&\times & \left[ A_k \, e^{i\vec{k}.\hat{z}}  \, \chi_k(t) +
A^{\dagger}_k \, e^{-i\vec{k}.\hat{z}}  \, \chi^*_k(t)\right],
\end{eqnarray}
where $\bar{e}_M^{AB}$ agrees with (\ref{pol}).

\section*{Final Comments}

We have studied the variational principle in presence of a
torsional geometry in presence of non-trivial boundary conditions. In the extended Einstein equations here obtained,
we have define an effective geometrically induced
energy momentum tensor for a Riemannian representation af a
torsional (Weitzenb\"ock) one. The energy-momentum tensor here obtained
must be viewed as representing geometrically induced matter from Weitzenb\"ock torsion. This
is the main difference with other approaches (for instance, the Space-Time-Matter theory\cite{STM}), where the energy-momentum tensor is induced from an extra dimensional
vacuum. However, in our approach the gravitational
wave dynamics on a Weitzenb\"{o}ck manifold the torsion and boundary terms must be taken into account in order to explain
the origin of the cosmological constant and GW. Our theory has many similitude to whole of Ferraro-Fiorini\cite{FF}. As it was shown in
(\ref{ca2}) the boundary terms could
be responsible for the cosmological constant which is produced by a source inside a 3D Gaussian (closed) hypersurface that
encloses that source. In the inflationary example studied in Sect. \ref{V} magnetic monopoles
are absent, due to the globally isotropy and homogeneity of the universe at large (cosmological) scales. This agrees with the absence of magnetic monopoles predicted by inflationary models.

\section{Acknowledgements}

\noindent J.E.Madriz Aguilar acknowledges CONACYT M\'exico, Centro Universitario de Ciencias Exactas e Ingenierias and Centro Universitario de los Valles, of Universidad de Guadalajara for financial support. J. M. Romero and M. Bellini acknowledge CONICET (PIP: 112-201101-00325) \& UNMdP (EXA651/14) for financial support.


\bigskip

\begin{thebibliography}{99}
\bibitem{HE}J. W. York, Phys. Rev. Lett. {\bf 16}: 1082 (1972);\\
G. W. Gibbons, S. W. Hawking, Phys. Rev. {\bf D10}: 2752 (1977).
\bibitem{2}L. S. Ridao; M. Bellini, Astrophys. Space Sci. {\bf 357}: 94 (2015).
\bibitem{1} J. Romero; E. Madriz Aguilar; M. Bellini, Phys.Dark Univ. {\bf 13}: 1 (2016).
\bibitem{3}O. Gr\"on, S. Hervik, {\em Einstein´s General Theory of Relativity}. Springer (2007).
\bibitem{4} M. Israelit, Gen. Rel. Grav. {\bf 29}: 1597 (1997); \\
M. Nakahara, {\em Geometry, Topology and Physics}, IoP, Bristol Philadephia, ISBN 0 7503 0606 8 (2003);\\
M. Israelit, Found. Phys. {\bf 35}: 1769 (2005); \\
H.I. Arcos, J.G. Pereira, Int.J.Mod.Phys. {\bf D13}: 2193 (2015).
\bibitem{5} H.Kleinert. {\em Gauge Fields in condensed matter}, p.1427, World Scientific,
Singapur (1989); \\
H. Kleinert. {\em Path Integrals in Quantum Mechanics, Statistics, Polymer Physics, and Financial Markets}.
p.786, 5th edition, pp. 1-1547, World Scientific, Singapore
(2009). Presented text is a free translation of the spanish
version.
\bibitem{6} E. Cartan, {\em Sur une généralisation de la notion de courbure de Riemann et les espaces à torsion}". C. R. Acad. Sci. (Paris) 174, 593–595 (1922);\\
E. Cartan, {\em Sur les variétés à connexion affine et la théorie de la relativité généralisée}. Part I: Ann. Éc. Norm. 40, 325–412
(1923) and ibid. 41, 1–25 (1924); Part II: ibid. 42, 17–88 (1925);\\
Qihong Huang, Puxun Wu, Hongwei Yu, Phys. Rev. {\bf D91}: 103502 (2015).
\bibitem{7} R. Weitzenb\"{o}ck. {\em Invarianten Theorie}. Noordhoff, Groningen, (1923);\\
The Weitzenb\"{o}ck geometry must be viewed as a particular case of the topics exposed in:
Arcos, H.I.; J.G. Pereira, Int. J. Mod. Phys. {\bf D13}: 2193 (2005).
\bibitem{8} J. Yepez, {\em Einstein’s vierbein field theory of curved space}. E-print arXiv: 1106.2037.
\bibitem{BD} T. S. Bunch and P. C. W. Davies, Proc. Roy.
Soc. {\bf A360}, 117 (1978).
\bibitem{STM} P. S. Wesson, Gen. Rel. Grav. {\bf 16}, 193 (1984);
P. Wesson, Gen. Rel. Grav. {\bf 22}, 707 (1990);
P. S. Wesson, Phys. Lett. {\bf B276}, 299 (1992);
P. S. Wesson and J. Ponce de Leon, J. Math. Phys. {\bf 33}, 3883 (1992);
H. Liu and P. S. Wesson, J. Math. Phys. {\bf 33}, 3888 (1992);
P. Wesson, H. Liu and P. Lim, Phys. Lett. {\bf B298}, 69 (1993).
\bibitem{FF} R. Ferraro, F. Friorini, Phys. Lett. {\bf B702}: 75 (2011).
\end{thebibliography}
\end{document}